\documentclass[onecolumn,trackchanges]{aastex6}

\fullcollaborationName{The Friends of AASTeX Collaboration}

\slugcomment{ApJ in press; received 2016 February 2; accepted 2016 April 7}
\shorttitle{Hot Jupiter Companions Inside $a_{\mathrm{ice}}$}
\shortauthors{Schlaufman \& Winn}

\begin{document}


\title{The Occurrence of Additional Giant Planets Inside the Water-Ice
Line in Systems with Hot Jupiters: Evidence Against High-Eccentricity
Migration}


\author{Kevin C. Schlaufman\altaffilmark{1,2,3,4,5,6} and Joshua N. Winn\altaffilmark{7,8}}
\affil{Kavli Institute for Astrophysics and Space Research,
Massachusetts Institute of Technology, Cambridge, MA 02139, USA}


\altaffiltext{1}{The Observatories of the Carnegie Institution
for Science, 813 Santa Barbara St., Pasadena, CA 91101, USA}
\altaffiltext{2}{Department of Astrophysical Sciences,
Princeton University, Princeton, NJ 08544, USA}
\altaffiltext{3}{Department of Physics and Astronomy, Johns Hopkins
University, Baltimore, MD 21218, USA}
\altaffiltext{4}{Kavli Fellow}
\altaffiltext{5}{Carnegie-Princeton Fellow}
\altaffiltext{6}{kcs@carnegiescience.edu}
\altaffiltext{7}{Department of Physics, Massachusetts Institute of
Technology, Cambridge, MA 02139, USA}
\altaffiltext{8}{jwinn@mit.edu}

\begin{abstract}

The origin of Jupiter-mass planets with orbital periods of only a few
days is still uncertain.  It is widely believed that these planets formed
near the water-ice line of the protoplanetary disk, and subsequently
migrated into much smaller orbits.  Most of the proposed migration
mechanisms can be classified either as disk-driven migration, or as
excitation of a very high eccentricity followed by tidal circularization.
In the latter scenario, the giant planet that is destined to become
a hot Jupiter spends billions of years on a highly-eccentric orbit,
with apastron near the water-ice line.  Eventually, tidal dissipation
at periastron shrinks and circularizes the orbit.  If this is correct,
then it should be especially rare for hot Jupiters to be accompanied by
another giant planet interior to the water-ice line.  Using the current
sample of giant planets discovered with the Doppler technique, we find
that hot Jupiters with $P_{\mathrm{orb}} < 10$ days are no more or less
likely to have exterior Jupiter-mass companions than longer-period giant
planets with $P_{\mathrm{orb}} \geq 10$ days.  This result holds for
exterior companions both inside and outside of the approximate location
of the water-ice line.  These results are difficult to reconcile with
the high-eccentricity migration scenario for hot Jupiter formation.

\end{abstract}

\keywords{planetary systems --- planets and satellites: detection ---
          planets and satellites: formation --- stars: statistics}



\section{Introduction}

Explaining the existence of hot Jupiters -- Jupiter-mass planets on
circular orbits with periods of only a few days -- is one of the oldest
problems in exoplanetary science.  Theoretical models suggested that while
it is formally possible for core accretion to produce a giant planet at
$a \lesssim 0.1$ AU around a Sun-like star, it is more likely that they
form beyond the location of the water-ice line in the protoplanetary
disk at $a \gtrsim 1$ AU and subsequently migrate into the close
proximity of the host star \citep{bod00}.  Several mechanisms for this
migration have been proposed.  In disk-driven migration, interactions
between a planet and the protoplanetary disk move a giant planet from
the water-ice line to the inner edge of the disk within the few Myr
of the disk's lifetime \citep[e.g.,][]{lin96}.  In high-eccentricity
migration, a giant planet initially forms on a circular orbit near
the water-ice line.  The orbital eccentricity is then excited to $e
\gtrsim 0.9$ in one of several possible ways: close encounters with other
planets, secular interactions with another massive body in the system,
or a combination of scattering and secular interactions.  Thereafter,
the strong tidal interactions between the planet and its host star at
periastron extract energy from the planetary orbit, leaving behind a
giant planet on a circular orbit with $P_{\mathrm{orb}} \lesssim 10$
days \citep[e.g.,][]{ras96,wei96,hol97,maz97,kis98,wu11,pet15}.

Several of the observed properties of the giant exoplanet
population support the high-eccentricity scenario. \citet{cum99} and
\citet{udr03} identified a ``three-day pile-up'' of hot Jupiters --
an enhancement in the occurrence rate $dN/d\log P_{\mathrm{orb}}$
at $P_{\mathrm{orb}} \approx 3$ days -- which is compatible with
the action of tidal circularization.  \citet{daw13} showed that this
pile-up is also apparent in the sample of metal-rich \textit{Kepler}
giant-planet-candidate host stars, although not among the more metal-poor
giant-planet-candidate hosts.  They argued that this is a natural outcome
of planet--planet scattering.  The reason is that metal-rich stars
likely had protoplanetary disks with more solid material and therefore
a better chance of forming a multiple-giant-planet system that could
undergo planet--planet scattering.  Protoplanetary disk around more
metal-poor stars likely had less solid material and therefore rarely
produced multiple-giant-planet systems capable of strong scattering.
In addition, the discovery of hot Jupiters for which the orbit is
significantly misaligned with the rotation of the host star is seemingly
at odds with disk migration, and naturally explained by high-eccentricity
migration \citep[e.g.,][]{fab07,jur08,cha08,nao11,nao12,wu11}.

Also widely cited as support for high-eccentricity migration is the idea
that hot Jupiters seem to be less likely than other types of planets
to be found with additional planetary companions.  In other words, hot
Jupiters have a reputation as ``lonely'' planets.  This inference relies
on a generic prediction of the high-eccentricity migration scenario: in
systems with a hot Jupiter, there should be few if any planets exterior
to the hot Jupiter and interior to the water-ice line.  The body that
gained the angular momentum lost by the hot-Jupiter progenitor must be
exterior to the water-ice line.  At the same time, any pre-existing
planets interior to the water-ice line would have been destabilized
during the high-eccentricity phase of the migration process, leading
to the ejection of that planet from that region or the interference
with the tidal circularization process.  Furthermore, some variants of
high-eccentricity migration rely on Kozai-Lidov oscillations excited by
a stellar companion, which are incapable of exciting high eccentricities
in the presence of companion planets \citep[e.g.,][]{koz62,lid62,wu03}.

In contrast to hot Jupiters, the population of ``warm Jupiters''
with $P_{\mathrm{orb}} \geq 10$ days probably did not form through
high-eccentricity migration, as star--planet tidal interactions at
their more distant periastron distances are not strong enough to shrink
their orbits.  Moreover, the possibility that warm Jupiters are actively
circularizing and are being observed in a low-eccentricity phase of
long-timescale Kozai-Lidov oscillations been investigated and found to
be in tension with results from Doppler surveys and the \textit{Kepler}
survey \citep{don14,daw15}.

The high-eccentricity scenario singles out hot Jupiters and predicts
they will have different companion statistics than warm Jupiters.
The frequency of long-period companions to hot Jupiters therefore
provides an observational test.  If the occurrence rate of long-period
companions to hot Jupiters inside the water-ice line were systematically
lower than the equivalent occurrence rate for cooler giant planets, it
would support the high-eccentricity migration scenario.  On the other
hand, if the occurrence rates were comparable, it would disfavor the
high-eccentricity migration scenario.  This work was motivated by the
fact that we were aware of the widely cited statement that hot Jupiters
are lonely, but could not find in the literature any quantitative results
for the conditional probability for a hot Jupiter to have a companion
inside the water-ice line.

Previous studies of the ``loneliness'' phenomenon have generally used
different metrics or had other features that prevented us from performing
the exact tests mentioned above.  \citet{wri09} found the distribution
of orbital distances of all the planets in multiple-planet systems to be
different from that of all the single-planet systems.  In particular,
the three-day pile-up discovered by \citet{cum99} was absent from
the multiple-planet systems.  The sample analyzed by \citet{wri09}
included systems discovered by both the Doppler and transit techniques.
Because the transit technique is especially biased toward short-period
planets, many of the apparently single systems analyzed by \citet{wri09}
may have long-period giant planet companions.

\citet{lat11} and \citet{ste12} found that short-period \textit{Kepler}
giant planet candidates are less likely to have small-planet candidate
companions than close-in Neptune-sized planet candidates or more distant
giant-planet candidates.  More than 50\% of the giant planet candidates
discovered by \textit{Kepler} have now been shown to be astrophysical
false positives \citep{san12,san16}.  The false-positive rate for
small-planet candidates appears to be much smaller \citep{fre13}.
An astrophysical false positive would be much less likely to present
evidence of additional planets than a genuine exoplanet system.

A direct approach to measuring the companion probability for close-in
giant planets is to undertake long-term Doppler observations of stars
known to have such planets.  Several groups have embarked on such efforts.
\citet{knu14} searched 48 systems with close-in Jupiter mass planets for
long-period companions, finding 14 systems likely to have an external
companion.  Assuming a double-power law for the companion mass--semimajor
axis distribution, they found that approximately 50\% of their targets
have a companion in the range 1 AU $\lesssim a \lesssim 20$ AU and
$1~M_{\mathrm{Jup}} \lesssim M_{p} \sin{i} \lesssim 13~M_{\mathrm{Jup}}$.
The same group has expanded their search for companions using direct
imaging to find distant companions and infrared spectroscopy to find
close companions \citep{ngo15,pis15}.  Most recently, \citet{bry16}
expanded their sample beyond hot Jupiter hosts to include giant
planets over a broader range of orbital distances.  They found that the
occurrence of companions in the range $1~M_{\mathrm{Jup}} \lesssim M_{p}
\sin{i} \lesssim 20~M_{\mathrm{Jup}}$ and 5 AU $\lesssim a \lesssim$
20 AU may be higher for hot Jupiters than for cooler giant planets.
That result is subject to the caveat that the companion statistics
may be systematically affected by stellar jitter (poorly understood,
time-correlated radial-velocity noise), the effect of which is magnified
in their hot Jupiter sample because of a relative lack of radial
velocity measurements.  They also did not make the specific comparison
between hot and cool Jupiters in the occurrence of companions within
the water-ice line.

In this paper, we compare the occurrence of long-period giant-planet
companions to both hot Jupiters and more distant giant planets.
We consider exterior companions both inside the water-ice line and out
to the completeness limit of our input sample.  We find that contrary
to the expectation from the high-eccentricity migration scenario,
there is no significant difference in the companion fraction for hot
Jupiters and more distant giant planets.  The same result applies to
companions both inside the water-ice line and out to the completeness
limit of our input data.  We describe our sample definition in Section 2,
we detail our analysis procedures in Section 3, we discuss the results
and implications in Section 4, and we summarize our findings in Section 5.

\section{Sample Definition}

We would like to calculate $P({\mathrm{LongPerJup}}|\mathrm{HotJup})$, the
conditional probability that a long-period Jupiter-mass planet exists in
a system given that it has a hot Jupiter. Rather than directly measuring
this quantity with long-term Doppler observations and high-contrast
imaging, we calculate it using Bayes' Theorem and publicly available data.
In particular, we use (i) the occurrence rates for giant planets as a
function of mass and orbital distance reported by \citet{cum08}; (ii)
the occurrence rate for hot Jupiters reported by \citet{wri12}; and
(iii) a sample of giant planets discovered by the Doppler technique.
We will refer to these data as the ``Doppler sample'' and describe its
construction below.

\begin{figure*}
\plottwo{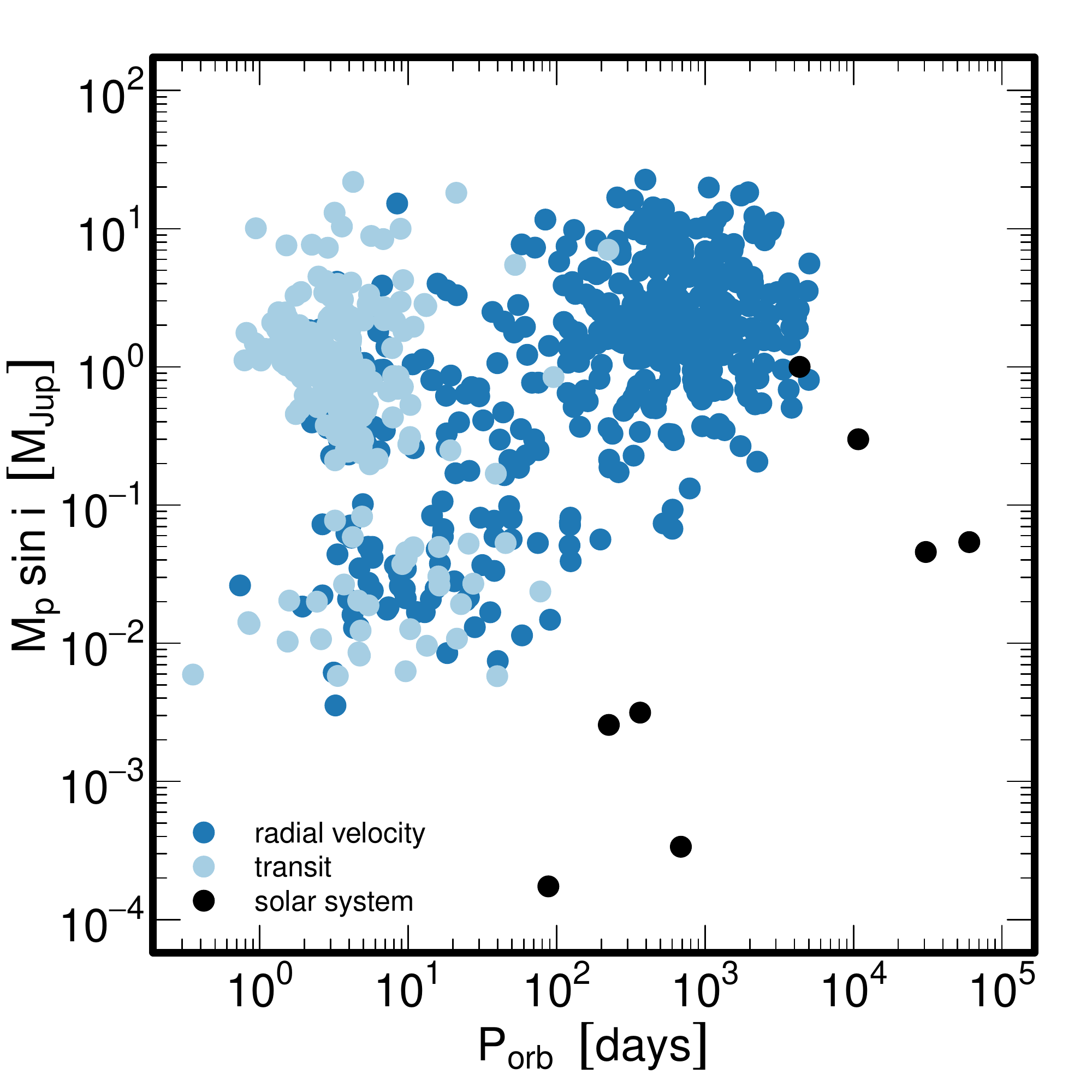}{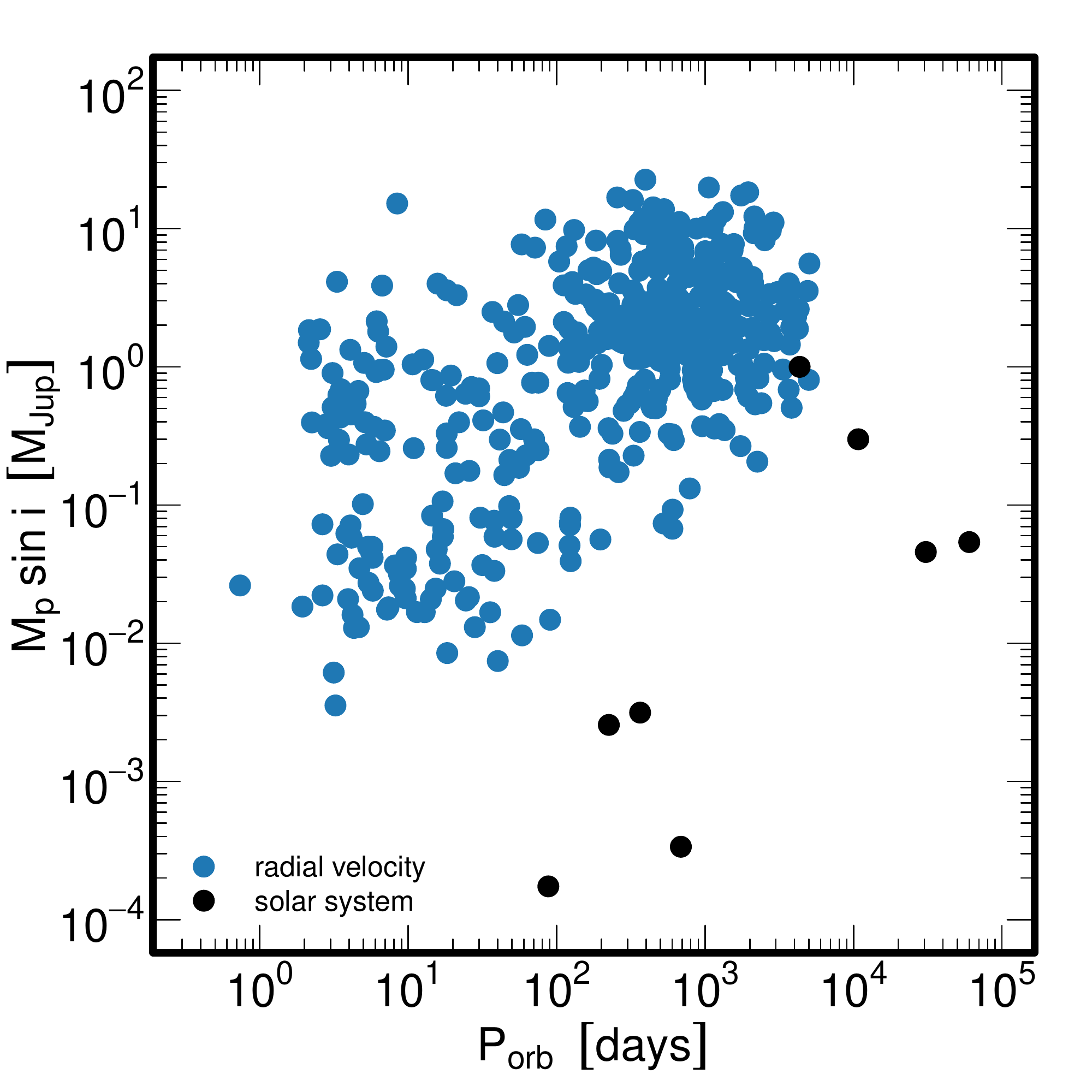}
\caption{Orbital period versus minimum planet mass for exoplanets and
solar system planets.  \textit{Left}: Including exoplanets discovered by
both the Doppler and transit techniques.  \textit{Right}: Only exoplanets
discovered by the Doppler technique.  We plot transit discoveries in
light blue, Doppler discoveries in dark blue, and solar system planets in
black.  Because transit surveys are extremely biased towards short-period
planets, the sample plotted on the left over-emphasizes the frequency
of hot Jupiters.\label{fig01}}
\end{figure*}

\begin{figure*}
\plottwo{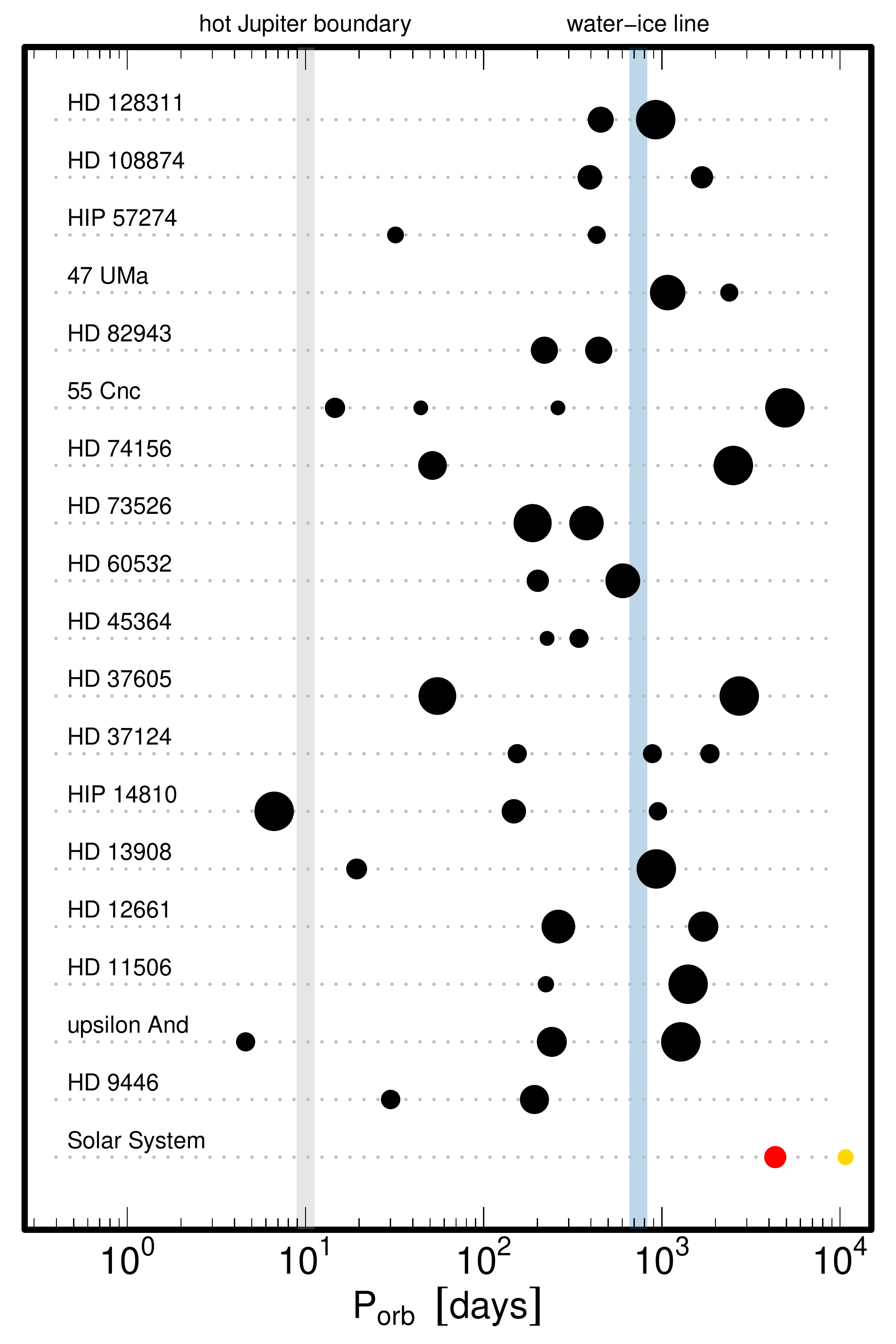}{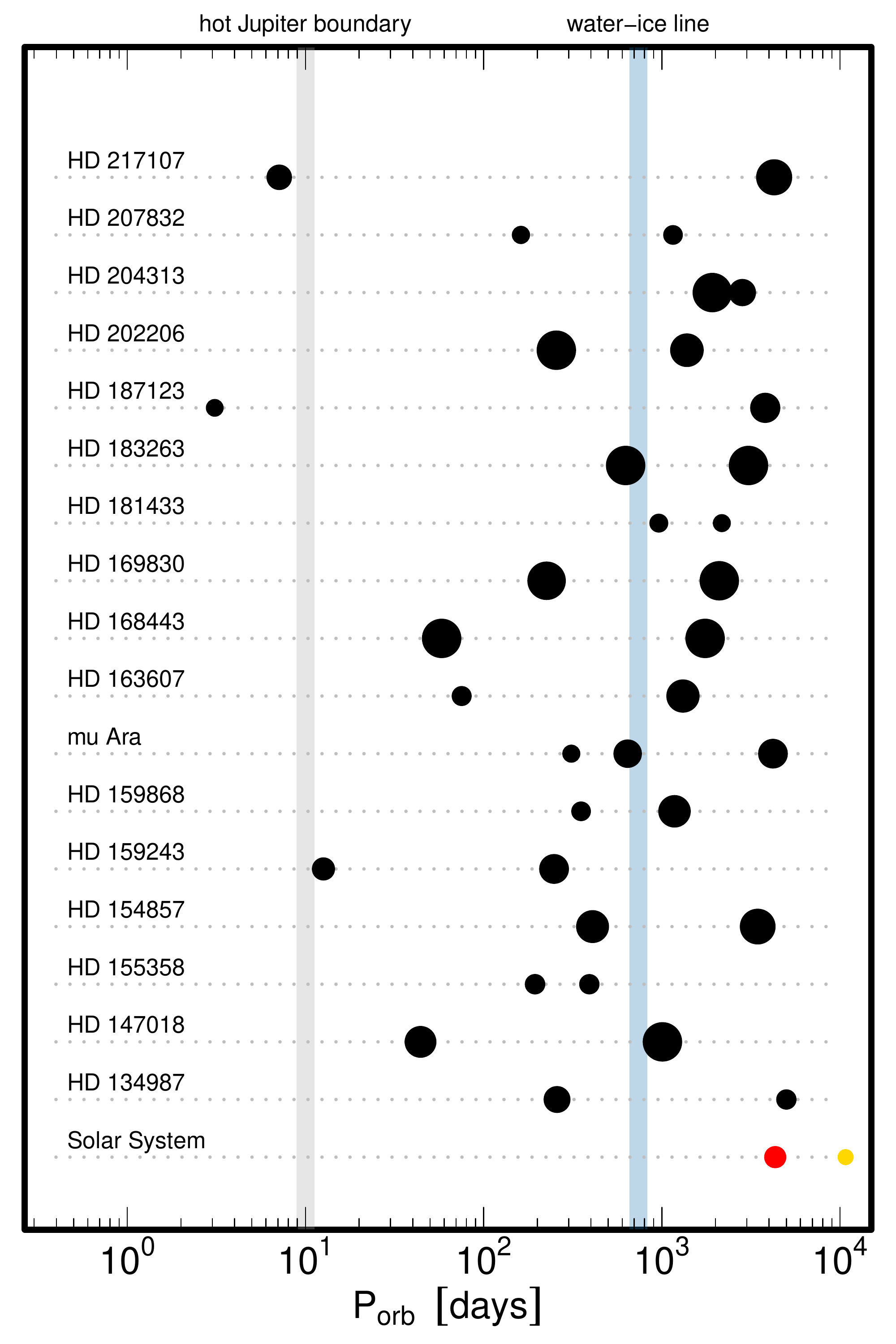}
\caption{Systems of multiple giant planets discovered with the radial
velocity technique around slightly evolved main-sequence FGK stars.
The size of a plotted point is proportional to $\log_{10}\!\left({M_{p}
\sin{i}}\right)$.  Four of the 35 multiple-giant-planet systems have a
hot Jupiter.\label{fig02}}
\end{figure*}

The planets discovered by the Doppler technique typically orbit bright
host stars that have been been subject to long-term high-precision
radial-velocity measurements, in some cases for more than 20 years.
This sample therefore provides the sensitivity to $M_{p} \sin{i} =
0.1~M_{\mathrm{Jup}}$ planets at $P_{\mathrm{orb}} \lesssim 100$ days
and to $M_{p} \sin{i} = 1~M_{\mathrm{Jup}}$ planets at $P_{\mathrm{orb}}
\lesssim 4080$ days \citep{cum08}.  We use \texttt{exoplanets.org} to
identify all giant exoplanets with $M_{p} \sin{i} > 0.1~M_{\mathrm{Jup}}$
discovered by the Doppler technique \citep{wri11,han14}.  We exclude
systems that were first discovered using the transit technique, because
of the extreme bias of the transit technique toward short-period planets
(Figure~\ref{fig01}).  We also wish to exclude evolved stars, because
they have markedly different giant-planet population statistics that
may be due to tidal evolution or other effects \citep[e.g.,][]{sch13}.
To identify and exclude evolved stars, we obtain for each star the
\textit{Hipparcos} parallaxes and Tycho $B-V$ colors from \citet{van07}
and apparent Tycho-2 $V$-band magnitudes from \citet{hog00}.  We then
select for main-sequence FGK stars showing little or no evolution,
using the criteria $\Delta M_{V} < 2.5$ and $B-V < 1.2$.  Here

\begin{eqnarray}
\Delta M_V & = & M_{V,MS}(B-V) - M_{V},
\end{eqnarray}

\noindent
and

\begin{eqnarray}
M_{V,MS}(B-V) & = & \sum_{i=0}^{9} a_{i} (B-V)^{i},
\end{eqnarray}

\noindent
where the coefficients $a_{i}$ are taken from \citet{wri05} and
represent the average \textit{Hipparcos} main sequence in the Tycho-2
system\footnote{The numerical values of the coefficients are $a_{i}$ =
(1.11255, 5.79062, $-16.76829$, 76.47777, $-140.08488$, 127.38044,
$-49.71805$, $-8.24265$, 14.07945, $-3.43155$).}.  Because of this
criterion we are forced to reject all planets orbiting stars without
\textit{Hipparcos} parallaxes.  We are left with 266 giant planets
orbiting 225 different stars and 35 multiple-giant-planet systems.
We list these systems in Table~\ref{tbl-1}.  The multiple-giant-planet
systems are the focus of this study, and we plot them in
Figure~\ref{fig02}.  We only use the Doppler sample to count interior
giant-planet companions to announced long-period giant planets.  Since the
radial velocity signal of interior giant planets always has a shorter
period and usually a larger amplitude than more distant giant planets,
the Doppler sample should be complete to such planets.

\begin{figure}
\plotone{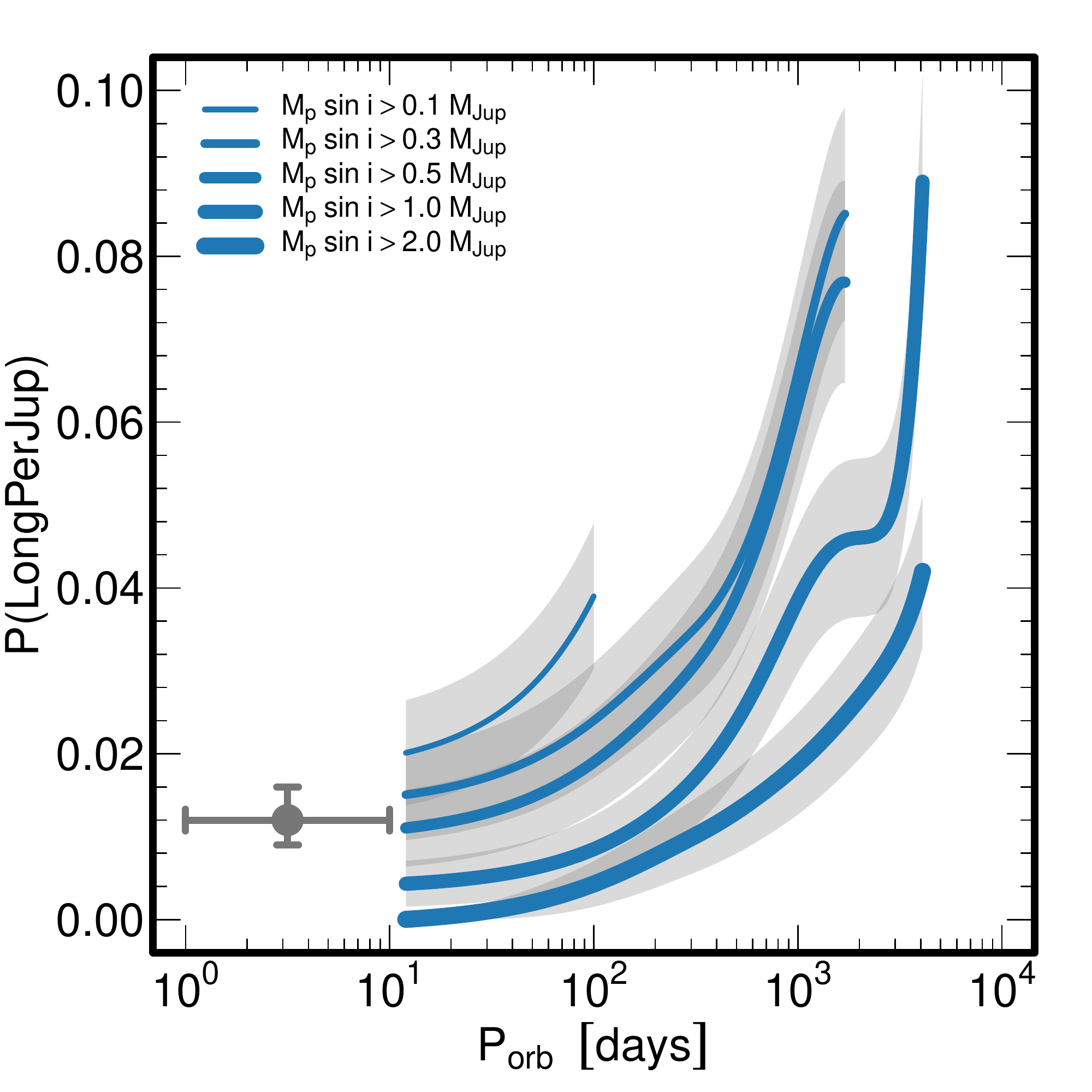}
\caption{Cumulative fraction of stars with a giant planet as a function
of period and $M_{p} \sin{i}$.  The hot Jupiter data point at $P < 10$
days is from \citet{wri12}, while the colored lines are from natural cubic
spline interpolation of the data in Table 1 of \citet{cum08} accounting
for completeness and including both announced and unannounced detections.
The gray regions are the 1-$\sigma$ uncertainties.\label{fig03}}
\end{figure}

We plot the giant planet occurrence from \citet{cum08} and
\citet{wri12} as a function of $P_{\mathrm{orb}}$ and $M_{p} \sin{i}$
in Figure~\ref{fig03}.  We assume that there is not a significant
undiscovered population of short-period giant planets in systems with
at least one long-period giant planet.  Given the biases of the Doppler
technique and the search strategy of existing Doppler surveys, this
should be a safe assumption.  In addition, we assume that the long-period
giant planet occurrence rate observed by \citet{cum08} applies to the
unobserved long-period giant planet population in the larger sample of
stars analyzed by \citet{wri12}.

\section{Analysis}

Let $P(\mathrm{HotJup})$ denote the probability that a star hosts
a hot Jupiter, as defined by \citet{wri12}: a planet with $M_{p}
\sin{i} > 0.1~M_{\mathrm{Jup}}$ and $P_{\mathrm{orb}} < 10$ days.
Let $P(\mathrm{LongPerJup})$ denote the probability that an FGK
star hosts a long-period Jovian planet, defined here as a planet
with $M_{p} \sin{i} > 1~M_{\mathrm{Jup}}$ and 10 days $\lesssim
P_{\mathrm{orb}} \lesssim 4080$ days.  The outer limit of 4080 days
is adopted to match that of \citet{cum08}, whose results are necessary
for our analysis.  Then by Bayes' Theorem, the conditional probability
$P(\mathrm{LongPerJup}|\mathrm{HotJup})$ that a system has a long-period
giant planet given that it has a hot Jupiter is

\begin{eqnarray}
P(\mathrm{LongPerJup}|\mathrm{HotJup}) & = &
\frac{P(\mathrm{HotJup}|\mathrm{LongPerJup}) P(\mathrm{LongPerJup})}
{P(\mathrm{HotJup}|\mathrm{LongPerJup}) P(\mathrm{LongPerJup}) +
 P(\mathrm{HotJup}|\mathrm{LongPerJup}') P(\mathrm{LongPerJup}')},
\end{eqnarray}

\noindent
where $P(\mathrm{HotJup}|\mathrm{LongPerJup})$ is the conditional
probability that a system with a long-period giant planet has a hot
Jupiter, $P(\mathrm{HotJup}|\mathrm{LongPerJup}')$ is the probability
that a system with no long-period giant planet has a hot Jupiter,
and $P(\mathrm{LongPerJup}')$ is the probability that a system has no
long-period giant planet.

First we will compute $P(\mathrm{HotJup}|\mathrm{LongPerJup})$.
The Doppler sample contains 136 systems with at least one long-period
giant planet, according the definition given above.  The Doppler sample
also includes four systems with both a hot Jupiter and at least one
long-period companion: $\upsilon$ And, HIP 14810, HD 187123, and HD
217107.  However, HD 217107c has an orbital period $P_{\mathrm{orb}} =
4270$ days, which is outside of the period range defined in Table 1 of
\citet{cum08} and cannot be used in this analysis.  Therefore, $P(HJ|GP)
= 3/136 = 0.020_{-0.010}^{+0.014}$.  We calculate our uncertainties
using a binomial distribution and an uninformative Beta$(\alpha,\beta$)
prior \citep[see Section 3.3 of][]{sch14}.

The hot Jupiter probability $P(\mathrm{HotJup})$ has already been the
subject of a careful study by \citet{wri12}.  They found 10 hot Jupiters
in their sample of 836 stars, using the criteria defined above, giving
$P(\mathrm{HotJup}) = 0.012_{-0.003}^{+0.004}$.

For $P(\mathrm{LongPerJup})$ we rely on the work by \citet{cum08}. They
found that 8.5\% of FGK stars have a $M_{p} \sin{i} = 1~M_{\mathrm{Jup}}$
planet in the range 11.5 days $< P_{\mathrm{orb}} < 4080$ days (or 0.1 AU
$< a < 5$ AU), giving $P(\mathrm{LongPerJup}) = 0.085_{-0.012}^{+0.013}$
and $P(\mathrm{LongPerJup}') = 1-P(\mathrm{LongPerJup}) =
0.915_{-0.012}^{+0.013}$.

To determine $P(\mathrm{HotJup}|\mathrm{LongPerJup}')$, we note that two
of the 10 hot Jupiters in the sample of \citet{wri12} have longer-period
companions.  None of the other eight hot Jupiters have longer-period
companions.  Although the sample of 836 stars from which they are drawn
has not been searched completely, we know from the occurrence rates
of \citet{cum08} that 8.5\% of them (on average) have longer-period
companions.  Therefore $P(\mathrm{HotJup}|\mathrm{LongPerJup}') =
(10-2)/\left[836(1-0.085)\right] = 0.010_{-0.003}^{+0.004}$.

We now have measurements and uncertainties for all the quantities on
the right side of Equation (3).  Plugging in the numbers reveals that
$P(\mathrm{LongPerJup}|\mathrm{HotJup}) = 0.155_{-0.077}^{+0.110}$.
It is interesting to compare this result to $P(\mathrm{LongPerJup}) =
0.085_{-0.012}^{+0.013}$.  Evidently, the fact that a hot Jupiter has
been observed in a system increases the probability that another giant
planet will be found within 5 AU in the same system, by nearly a factor
of two (with a large statistical uncertainty).

To test the prediction of the high-eccentricity migration scenario,
we need to calculate $P(\mathrm{LongPerJup}|\mathrm{HotJup})$ for
the specific case of companions that are inside the water-ice line.
\citet{mul15} determined that the water-ice line in a solar-composition
protoplanetary disk around a Sun-like star during the planet formation
epoch is at $a \approx 1.6$ AU, or $P_{\mathrm{orb}} \approx 739$ days.
Hence we define a long-period Jovian planet inside the water-ice line
(InsideIceJup) as a planet with $M_{p} \sin{i} > 0.3~M_{\mathrm{Jup}}$ and
10 days $< P < 739$ days, and repeat the calculation from the preceding
paragraph.  Using this revised definition, only the $\upsilon$~And and
HIP~14810 systems have both a hot Jupiter and at least one long-period
companion, and there are 105 systems with at least one long-period
giant planet. Therefore, $P(\mathrm{HotJup}|\mathrm{InsideIceJup}) =
2/105 = 0.016_{-0.009}^{+0.016}$.  The value of $P(\mathrm{HotJup}) =
0.012_{-0.003}^{+0.004}$ is the same as before.  The overall giant-planet
occurrence rate from \citet{cum08} is now 3.9\% for FGK stars in the
range 11.5 days $< P_{\mathrm{orb}} < 739$ days (or 0.1 AU $< a < 1.6$
AU), giving $P(\mathrm{InsideIceJup}) = 0.039_{-0.008}^{+0.009}$
and $P(\mathrm{InsideIceJup}') = 1-P(\mathrm{InsideIceJup}) =
0.961_{-0.008}^{+0.009}$.  Only $\upsilon$ And b is in the sample
of \citet{wri12}, so $P(\mathrm{HotJup}|\mathrm{InsideIceJup}) =
(10-1)/\left[836(1-0.039)\right] = 0.011_{-0.004}^{+0.004}$.  Once again
using Equation (3) with InsideIceJup in place of LongPerJup, we find
that $P(\mathrm{InsideIceJup}|\mathrm{HotJup}) = 0.055_{-0.033}^{+0.061}$.

To complete the test of the high-eccentricity migration scenario, we need
to compare $P(\mathrm{LongPerJup}|\mathrm{HotJup})$ for hot Jupiters
to $P[\mathrm{LongPerJup}|\mathrm{Jup(}{P_{\mathrm{orb}}}\mathrm{)}]$,
where $P[\mathrm{LongPerJup}|\mathrm{Jup(}{P_{\mathrm{orb}}}\mathrm{)}]$
is the occurrence of long-period Jovian companions to Jovian planets
at an arbitrary period $P_{\mathrm{orb}}$.  In this way we can evaluate
$P[\mathrm{LongPerJup}|\mathrm{Jup(}{P_{\mathrm{orb}}}\mathrm{)}]$ for
warm Jupiters and compare it to the corresponding conditional probability
for hot Jupiters.  Because the sample in \citet{wri12} is specific
to hot Jupiters, in the general calculation we only use the sample of
multiple-giant-planet systems described in Section 2 and the data in Table
1 of \citet{cum08}.  We define an arbitrary period $P_{\mathrm{cut}}$
that separates warm giant planets from their longer-period companions.
We also define a minimum mass $M_{\mathrm{in}}$ for giant planets inside
of $P_{\mathrm{cut}}$ and a minimum mass $M_{\mathrm{out}}$ for giant
planets outside of $P_{\mathrm{cut}}$.

Using the exoplanet sample from \texttt{exoplanets.org}, first we
calculate $N_{\mathrm{comp}}$, the number of multiple-planet systems with
at least one component with $M > M_{\mathrm{in}}$ and $P_{\mathrm{orb}} <
P_{\mathrm{cut}}$ as well as a long-period companion at $P_{\mathrm{cut}}
< P_{\mathrm{orb}} < P_{\mathrm{out}}$, where $P_{\mathrm{out}}$ is the
$M_{\mathrm{out}}$-dependent completeness limit from \citet{cum08}.
We next calculate $N_{\mathrm{in}}$, the number of systems with at
least one component with $M > M_{\mathrm{in}}$ and $P_{\mathrm{orb}} <
P_{\mathrm{cut}}$.  Finally, we calculate $N_{\mathrm{out}}$, the number
of systems with at least one component with $M > M_{\mathrm{out}}$
and $P_{\mathrm{cut}} < P_{\mathrm{orb}} < P_{\mathrm{out}}$.
We calculate $P[\mathrm{Jup(}{P_{\mathrm{orb}}}\mathrm{)}]$ and
$P(\mathrm{LongPerJup})$ -- the giant planet occurrence rates
inside and outside of $P_{\mathrm{cut}}$ -- by two-dimensional
interpolation of Table 1 from \citet{cum08}.  By the axioms of
probability, $P(\mathrm{LongPerJup}') = 1 - P(\mathrm{LongPerJup})$
and $P[\mathrm{Jup(}{P_{\mathrm{orb}}}\mathrm{)}|\mathrm{LongPerJup}]
= N_{\mathrm{comp}}/N_{\mathrm{out}}$.  The effective size of our
multiple-giant-planet sample is $N_{\mathrm{out}}/P(\mathrm{LongPerJup})$,
implying

\begin{eqnarray}
P(\mathrm{LongPerJup}') = \frac{N_{\mathrm{in}} - N_{\mathrm{comp}}}
{N_{\mathrm{out}} \left[1/P(\mathrm{LongPerJup}) - 1\right] }.
\end{eqnarray}

The probability that a warm Jupiter has a long-period companion
$P[\mathrm{LongPerJup}|\mathrm{Jup(}{P_{\mathrm{orb}}}\mathrm{)}]$ is then

\begin{eqnarray}
P[\mathrm{LongPerJup}|\mathrm{Jup(}{P_{\mathrm{orb}}}\mathrm{)}] & = &
\frac{P[\mathrm{Jup(}{P_{\mathrm{orb}}}\mathrm{)}|\mathrm{LongPerJup}] P(\mathrm{LongPerJup})}
{P[\mathrm{Jup(}{P_{\mathrm{orb}}}\mathrm{)}|\mathrm{LongPerJup}] P(\mathrm{LongPerJup}) +
P[\mathrm{Jup(}{P_{\mathrm{orb}}}\mathrm{)}|\mathrm{LongPerJup}'] P(\mathrm{LongPerJup}')}.
\nonumber\\
\end{eqnarray}

\begin{figure*}
\plottwo{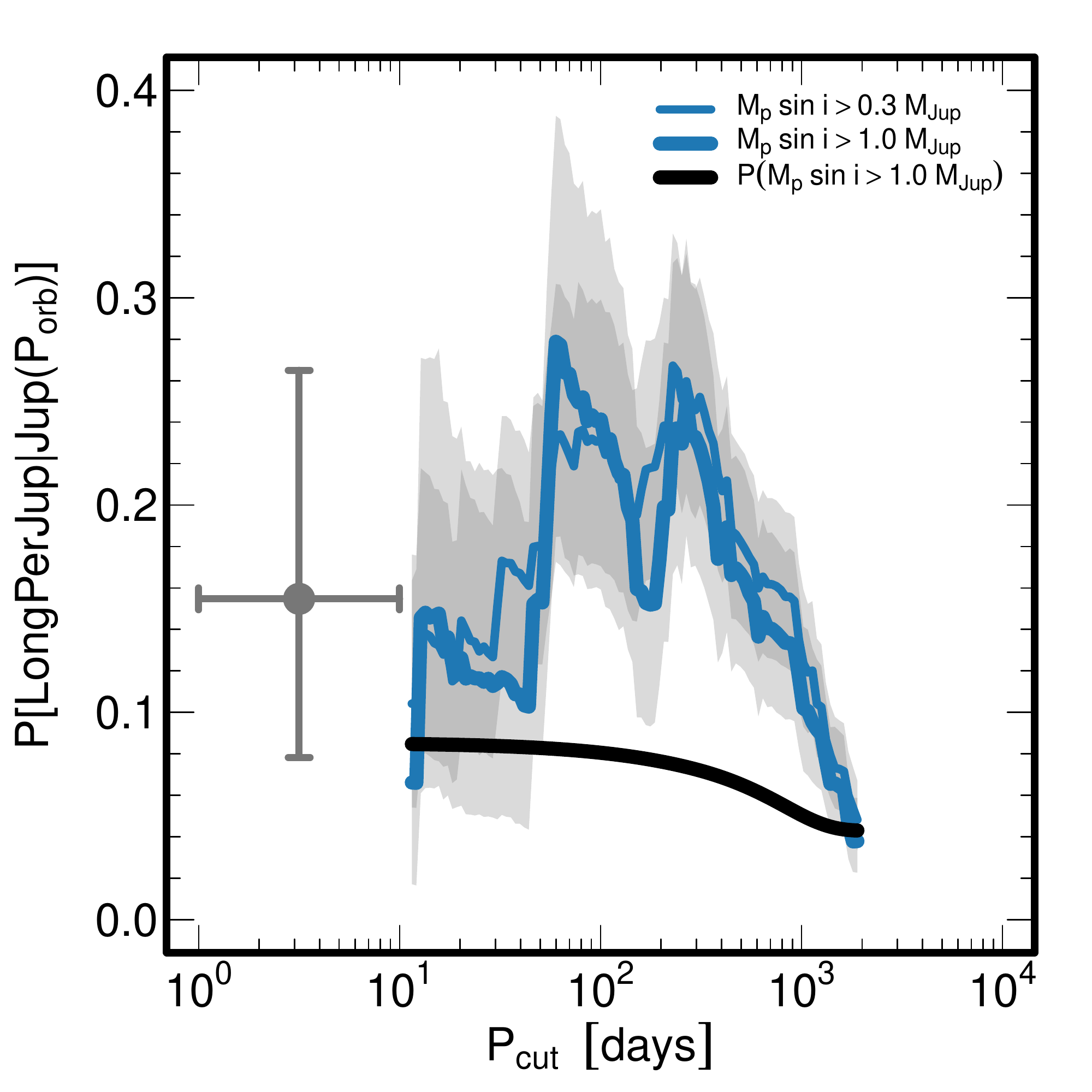}{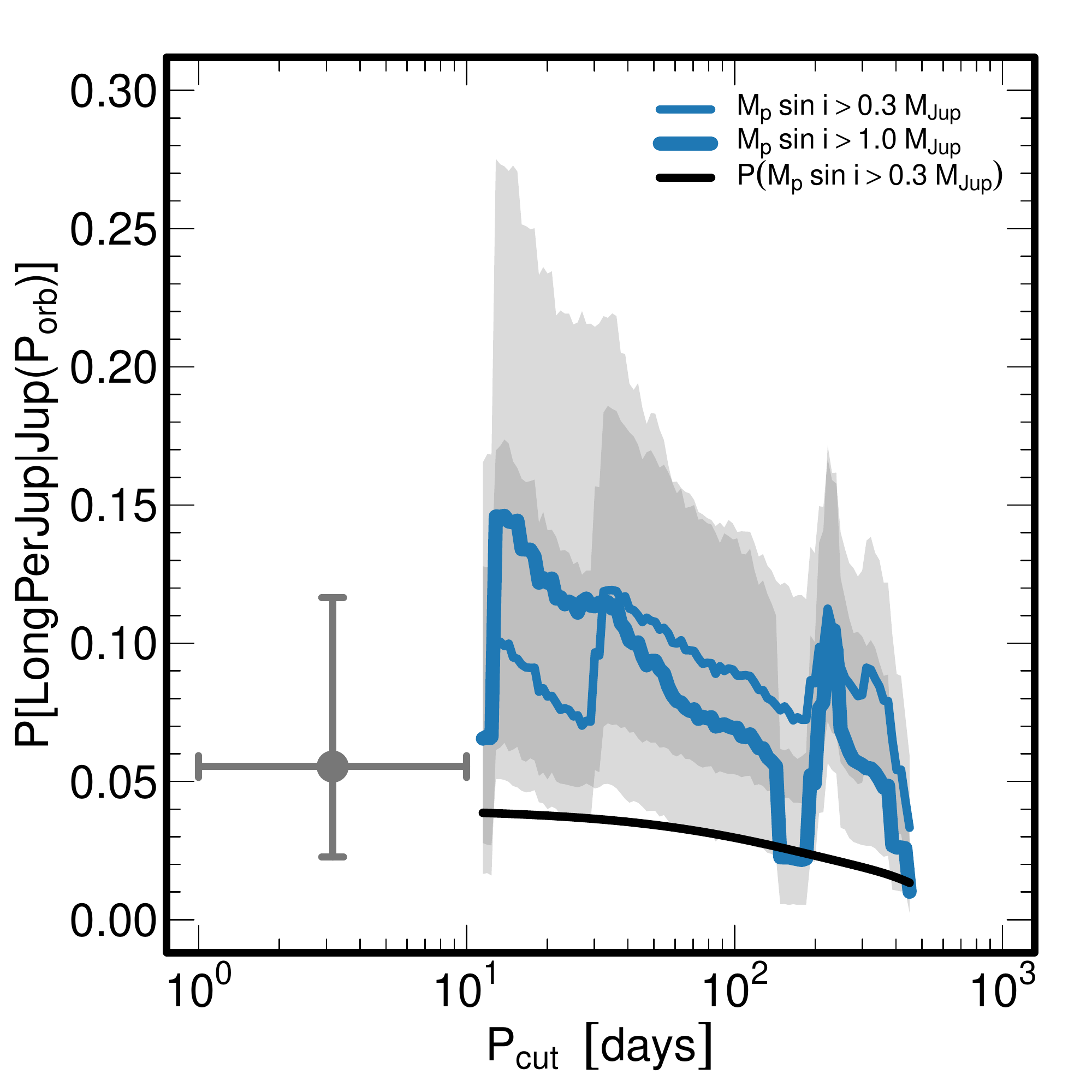}
\caption{\textit{Left}: Conditional probability of the presence
of an exterior, long-period giant planet with $M_{p} \sin{i} >
1~M_{\mathrm{Jup}}$ and $P_{\mathrm{orb}} < 4080$ days given the
presence of an interior giant planet with period and minimum mass
as indicated.  The gray regions are the 1-$\sigma$ uncertainties and
the solid black curve is the probability $P(\mathrm{LongPerJup})$.
\textit{Right}: Conditional probability of the presence of an
exterior, long-period giant planet inside the water-ice line with
$M_{p} \sin{i} > 0.3~M_{\mathrm{Jup}}$ and $P_{\mathrm{orb}} < 739$
days given the presence of an interior giant planet with period
and minimum mass as indicated.  We assume a water-ice line at $a =
1.6$ AU, or $P_{\mathrm{orb}} = 739$ days for a one solar mass star
\citep{mul15}.  The gray regions are the 1-$\sigma$ uncertainties and
the solid black curve is the probability $P(\mathrm{LongPerJup})$.
The apparent scatter at short orbital period is smaller than the
measurement uncertainty, so there is no evidence for a dependence of
$P[\mathrm{LongPerJup}|\mathrm{Jup(}{P_{\mathrm{orb}}}\mathrm{)}]$
on minimum mass.  The trend towards smaller
$P[\mathrm{LongPerJup}|\mathrm{Jup(}{P_{\mathrm{orb}}}\mathrm{)}]$ as
$P_{\mathrm{cut}}$ increases is statistically real, but best explained by
the decreasing range in orbital period available to external companions
between the interior giant planet and the completeness limit as the period
of the inner planet increases.  The fact that hot Jupiters are just as
likely as more distant giant planets to have long-period companions
inside the water-ice line is contrary to the expectation from simple
models of high-eccentricity migration.\label{fig04}}
\end{figure*}

\noindent
Figure~\ref{fig04} shows a comparison between the hot Jupiter and warm
Jupiter companion probabilities as a function of $M_{\mathrm{in}}$
and $P_{\mathrm{cut}}$.  There is no evidence that hot Jupiters are less
likely to have long-period giant-planet companions than cooler Jupiters,
either inside the water-ice line or out to 5 AU.  We give sample values
in Table~\ref{tbl-2} and Table~\ref{tbl-3}.

\begin{deluxetable*}{llll}
\tablecaption{Occurrence of Long-period Companions with $a < 5$ AU to Short-period Giant Planets\label{tbl-2}}
\tablewidth{0pt}
\tablehead{\colhead{Probability} &
\colhead{Hot Jupiter\tablenotemark{1}} &
\colhead{Warm Jupiter\tablenotemark{2}} &
\colhead{Mild Jupiter\tablenotemark{3}}}
\startdata
$P[\mathrm{Jup(}{P_{\mathrm{orb}}}\mathrm{)}]$        & $0.012_{-0.003}^{+0.004}$ & $0.038_{-0.008}^{+0.010}$ & $0.036_{-0.008}^{+0.009}$ \\
$P(\mathrm{LongPerJup})$       & $0.085_{-0.012}^{+0.013}$ & $0.080_{-0.012}^{+0.013}$ & $0.070_{-0.011}^{+0.012}$ \\
$P(\mathrm{LongPerJup}')$   & $0.915_{-0.012}^{+0.013}$ & $0.920_{-0.013}^{+0.012}$ & $0.930_{-0.012}^{+0.011}$ \\
$P[\mathrm{Jup(}{P_{\mathrm{orb}}}\mathrm{)}|\mathrm{LongPerJup}]$     & $0.020_{-0.010}^{+0.014}$ & $0.100_{-0.024}^{+0.029}$ & $0.213_{-0.038}^{+0.043}$ \\
$P[\mathrm{Jup(}{P_{\mathrm{orb}}}\mathrm{)}|\mathrm{LongPerJup}']$ & $0.010_{-0.003}^{+0.004}$ & $0.036_{-0.005}^{+0.005}$ & $0.053_{-0.006}^{+0.006}$ \\
$P[\mathrm{LongPerJup}|\mathrm{Jup(}{P_{\mathrm{orb}}}\mathrm{)}]$     & $0.155_{-0.077}^{+0.110}$ & $0.195_{-0.049}^{+0.057}$ & $0.231_{-0.048}^{+0.054}$
\enddata
\tablenotetext{1}{Defined as a planet with $M_{p} \sin{i} >
0.1~M_{\mathrm{Jup}}$ and $P_{\mathrm{orb}} < 10$ days.}
\tablenotetext{2}{Defined as a planet with $M_{p} \sin{i} >
0.1~M_{\mathrm{Jup}}$ and 10 days $< P_{\mathrm{orb}} < 100$ days.}
\tablenotetext{3}{Defined as a planet with $M_{p} \sin{i} >
0.3~M_{\mathrm{Jup}}$ and 10 $< P_{\mathrm{orb}} < 365$ days.}
\end{deluxetable*}

\begin{deluxetable*}{llll}
\tablecaption{Occurrence of Long-period Companions with $a < 1.6$ AU to Short-period Giant Planets\label{tbl-3}}
\tablewidth{0pt}
\tablehead{\colhead{Probability} &
\colhead{Hot Jupiter\tablenotemark{1}} &
\colhead{Warm Jupiter\tablenotemark{2}} &
\colhead{Mild Jupiter\tablenotemark{3}}}
\startdata
$P[\mathrm{Jup(}{P_{\mathrm{orb}}}\mathrm{)}]$        & $0.012_{-0.003}^{+0.004}$ & $0.039_{-0.008}^{+0.009}$ & $0.036_{-0.008}^{+0.009}$ \\
$P(\mathrm{LongPerJup})$       & $0.039_{-0.008}^{+0.009}$ & $0.029_{-0.007}^{+0.008}$ & $0.016_{-0.005}^{+0.007}$ \\
$P(\mathrm{LongPerJup}')$   & $0.961_{-0.008}^{+0.009}$ & $0.971_{-0.008}^{+0.007}$ & $0.984_{-0.007}^{+0.005}$ \\
$P[\mathrm{Jup(}{P_{\mathrm{orb}}}\mathrm{)}|\mathrm{LongPerJup}]$     & $0.016_{-0.009}^{+0.016}$ & $0.070_{-0.025}^{+0.032}$ & $0.195_{-0.058}^{+0.069}$ \\
$P[\mathrm{Jup(}{P_{\mathrm{orb}}}\mathrm{)}|\mathrm{LongPerJup'}]$ & $0.011_{-0.004}^{+0.004}$ & $0.022_{-0.003}^{+0.003}$ & $0.037_{-0.004}^{+0.004}$ \\
$P[\mathrm{LongPerJup}|\mathrm{Jup(}{P_{\mathrm{orb}}}\mathrm{)}]$     & $0.055_{-0.033}^{+0.061}$ & $0.086_{-0.034}^{+0.046}$ & $0.078_{-0.031}^{+0.042}$
\enddata
\tablenotetext{1}{Defined as a planet with $M_{p} \sin{i} >
0.1~M_{\mathrm{Jup}}$ and $P_{\mathrm{orb}} < 10$ days.}
\tablenotetext{2}{Defined as a planet with $M_{p} \sin{i} >
0.1~M_{\mathrm{Jup}}$ and 10 days $< P_{\mathrm{orb}} < 100$ days.}
\tablenotetext{3}{Defined as a planet with $M_{p} \sin{i} >
0.3~M_{\mathrm{Jup}}$ and 10 $< P_{\mathrm{orb}} < 365$ days.}
\end{deluxetable*}

\section{Discussion}

We have shown that hot Jupiters are just as likely as cooler Jupiters
to have long-period Jupiter-mass companions.  This observation
applies both for companions inside the water-ice line and out to the
completeness limit of long-term Doppler surveys.  Our estimate of the
fraction of warm Jupiters that have long-period giant-planet companions
$P[\mathrm{LongPerJup}|\mathrm{Jup(}{P_{\mathrm{orb}}}\mathrm{)}] \approx
0.2$ is consistent with the multiplicity rate determined by previous
studies \citep[e.g.,][]{wri07,wri09}.  These results are inconsistent
with the expectation from the simplest models of high-eccentricity
migration that suggest that most giant planets form beyond the water-ice
line, lose angular momentum to another body in the system, have their
eccentricities excited to $e \gtrsim 0.9$, lose orbital energy to tidal
decay, and circularize in the close proximity of their host stars.

How can one reconcile these findings with the previous work by
\citet{lat11} and \citet{ste12}, which found that giant-planet sized
\textit{Kepler} planet candidates are less likely to have additional
transiting planets in the system or measurable transit-timing variations
(TTVs) than smaller or more distant \textit{Kepler} planet candidates?
One possibility is that many of the giant-planet sized \textit{Kepler}
planet candidates considered by \citet{lat11} and \citet{ste12} may
have been false positives.  \citet{san12,san16} have shown that more
than 50\% of the giant-planet sized \textit{Kepler} planet candidates
are false positives of some kind, and false positives are not likely to
show evidence for additional transiting planets or TTVs.  More generally,
those previous studies were based on samples of transiting planets that
are more strongly biased toward close-in planets and companions.

Many of the other observations usually cited in support of
high-eccentricity migration are also being reexamined.  The reality of
the three-day pile-up of hot Jupiters, first identified in \citet{cum99},
seemed questionable in light of the \textit{Kepler} giant planet candidate
discoveries \citep[e.g.,][]{you11,how12,fre13}.  More recently, the
existence of a peak in the period distribution was identified in a sample
of confirmed \textit{Kepler} giant planets through the groundbreaking
work of \citet{san12,san16}. However the observations are compatible
with a rather broad peak, over the range 1 day $\lesssim P_{\mathrm{orb}}
\lesssim 10$ days, and a narrower period valley than had been previously
thought.  A broad pile-up and a narrow period valley would be difficult
to reconcile with high-eccentricity migration and tidal circularization,
as those models predict a sharp peak at twice the tidal radius $r_{t}
= R_{p} (M_{\ast}/M_{p})^{1/3}$ followed by a broad period valley.
The broader three-day pile-up can be accommodated in the disk migration
scenario \citep[e.g.,][]{ida04}.

The existence of significant spin--orbit misalignment can
also be explained without high-eccentricity migration.  Some
candidates mechanisms are star formation itself, hydrodynamic
disk-planet or magnetic star-disk interactions, torques from
distant stellar companions, or even stochastic rearrangement
of angular momentum mediated by gravity waves within hot stars
\citep[e.g.,][]{bat10,thi11,fie15,tre91,tho03,lai11,fou11,inn97,bat12,sto14,rog12}.

While we have argued that our measurement is inconsistent with the
expectation from simple models of high-eccentricity migration, our result
is consistent with models such as those proposed by \citet{gui11} that
require significant disk migration before eccentricity excitation.
It is also consistent with both the classical disk migration
scenario and the current generation of in-situ formation models
\citep[e.g.,][]{lin96,bol16,bat15}.

\section{Conclusions}

We found that hot Jupiters are just as likely as cooler giant planets
to have long-period Jupiter-mass companions.  This results applies
to long-period giant-planet companions both inside and outside the
water-ice line.  This observation is not expected if hot Jupiters are
produced by high-eccentricity migration and therefore emphasizes the
importance of either disk-driven migration or in situ formation for the
existence of short-period giant planets.

\acknowledgments
We thank Konstantin Batygin, Kat Deck, and Cristobal Petrovich for
helpful discussions.  This research has made use of NASA's Astrophysics
Data System Bibliographic Services, the SIMBAD database, operated at CDS,
Strasbourg, France \citep{wen00}, as well as the Exoplanet Orbit Database
and the Exoplanet Data Explorer at exoplanets.org.  Support for this
work was provided by the MIT Kavli Institute for Astrophysics and Space
Research through a Kavli Postdoctoral Fellowship and by the Carnegie
Institution for Science through a Carnegie-Princeton Postdoctoral
Fellowship.

\begin{deluxetable}{lc}
\tablecaption{Exoplanet Systems in Our Doppler Sample\label{tbl-1}}
\tablewidth{0pt}
\tablehead{\colhead{System} & \colhead{$N$\tablenotemark{1}}}
\startdata
HD 142 & 1 \\
HD 1237 & 1 \\
HD 1666 & 1 \\
HD 2039 & 1 \\
HIP 2247 & 1 \\
HD 2638 & 1 \\
HD 3651 & 1 \\
HD 4113 & 1 \\
HD 4208 & 1 \\
HD 4203 & 1 \\
HD 5388 & 1 \\
HD 6434 & 1 \\
HIP 5158 & 1 \\
HD 6718 & 1 \\
HD 7199 & 1 \\
HD 7449 & 1 \\
HD 8535 & 1 \\
HD 8574 & 1 \\
HD 9446 & 2 \\
ups And & 3 \\
HD 10180 & 1 \\
HD 10647 & 1 \\
HD 10697 & 1 \\
HD 11506 & 2 \\
HD 11964 & 1 \\
HD 12661 & 2 \\
GJ 86 & 1 \\
HD 13931 & 1 \\
HD 13908 & 2 \\
HD 16141 & 1 \\
30 Ari B & 1 \\
HD 16175 & 1 \\
HD 16760 & 1 \\
iot Hor & 1 \\
HD 17156 & 1 \\
HIP 14810 & 3 \\
HD 19994 & 1 \\
HD 20782 & 1 \\
HD 20868 & 1 \\
eps Eri & 1 \\
HD 23127 & 1 \\
HD 23079 & 1 \\
HD 22781 & 1 \\
HD 23596 & 1 \\
HD 24040 & 1 \\
HD 25171 & 1 \\
HD 27894 & 1 \\
HD 28254 & 1 \\
HD 28185 & 1 \\
HD 30177 & 1 \\
HD 30562 & 1 \\
HD 31253 & 1 \\
HD 33283 & 1 \\
HD 33636 & 1 \\
HD 34445 & 1 \\
HD 33564 & 1 \\
HD 290327 & 1 \\
HD 38283 & 1 \\
HD 37124 & 3 \\
HD 39091 & 1 \\
HD 37605 & 2 \\
HD 40979 & 1 \\
HD 43197 & 1 \\
HD 43691 & 1 \\
HD 44219 & 1 \\
HD 45364 & 2 \\
HD 45350 & 1 \\
HD 45652 & 1 \\
HD 46375 & 1 \\
HD 47186 & 1 \\
HD 48265 & 1 \\
HD 49674 & 1 \\
HD 50499 & 1 \\
HD 50554 & 1 \\
HD 52265 & 1 \\
HD 60532 & 2 \\
HD 63454 & 1 \\
HD 63765 & 1 \\
HD 65216 & 1 \\
HD 66428 & 1 \\
HD 67087 & 1 \\
HD 68988 & 1 \\
HD 70642 & 1 \\
HD 72659 & 1 \\
HD 73267 & 1 \\
HD 73256 & 1 \\
HD 73526 & 2 \\
HD 74156 & 2 \\
HD 75289 & 1 \\
55 Cnc & 4 \\
HD 75898 & 1 \\
HD 76700 & 1 \\
HD 79498 & 1 \\
HD 81040 & 1 \\
HD 82943 & 2 \\
HD 83443 & 1 \\
HD 85390 & 1 \\
HD 86081 & 1 \\
HD 86264 & 1 \\
BD-08 2823 & 1 \\
HD 87883 & 1 \\
HD 89307 & 1 \\
HD 89744 & 1 \\
HD 92788 & 1 \\
HD 93083 & 1 \\
47 UMa & 2 \\
HD 96167 & 1 \\
HD 99109 & 1 \\
HD 99492 & 1 \\
HD 100777 & 1 \\
HD 101930 & 1 \\
HIP 57274 & 2 \\
HD 102117 & 1 \\
HD 102195 & 1 \\
HD 103774 & 1 \\
HD 104067 & 1 \\
HD 106252 & 1 \\
HD 107148 & 1 \\
HD 108147 & 1 \\
HD 108341 & 1 \\
HD 108874 & 2 \\
HD 109246 & 1 \\
HD 109749 & 1 \\
HD 111232 & 1 \\
HD 113337 & 1 \\
HD 114386 & 1 \\
HD 114613 & 1 \\
HD 114762 & 1 \\
HD 114783 & 1 \\
HD 114729 & 1 \\
70 Vir & 1 \\
HD 117207 & 1 \\
HD 117618 & 1 \\
HD 118203 & 1 \\
tau Boo & 1 \\
HD 121504 & 1 \\
HD 125612 & 1 \\
HD 126614A & 1 \\
HD 128311 & 2 \\
HD 130322 & 1 \\
HD 132406 & 1 \\
HD 132563B & 1 \\
HD 131664 & 1 \\
HD 134987 & 2 \\
HD 136118 & 1 \\
HD 137388 & 1 \\
HD 330075 & 1 \\
HD 141937 & 1 \\
HD 142415 & 1 \\
rho CrB & 1 \\
HD 143361 & 1 \\
HD 142022 & 1 \\
14 Her & 1 \\
HD 145377 & 1 \\
HD 147018 & 2 \\
HD 147513 & 1 \\
HD 148156 & 1 \\
HD 149026 & 1 \\
HD 149143 & 1 \\
HD 154345 & 1 \\
HD 153950 & 1 \\
HD 155358 & 2 \\
HD 154672 & 1 \\
HD 154857 & 2 \\
HD 156279 & 1 \\
HD 156411 & 1 \\
HD 156846 & 1 \\
HD 159243 & 2 \\
HD 159868 & 2 \\
mu Ara & 3 \\
HD 162020 & 1 \\
HD 163607 & 2 \\
HD 164509 & 1 \\
HD 164922 & 1 \\
HD 164604 & 1 \\
HD 168443 & 2 \\
HD 168746 & 1 \\
HD 169830 & 2 \\
HD 170469 & 1 \\
HD 171238 & 1 \\
HD 175167 & 1 \\
HD 178911B & 1 \\
HD 179949 & 1 \\
HD 181720 & 1 \\
HD 181433 & 2 \\
HD 183263 & 2 \\
HD 231701 & 1 \\
HD 185269 & 1 \\
16 Cyg B & 1 \\
HD 187123 & 2 \\
HD 187085 & 1 \\
HD 188015 & 1 \\
HD 189733 & 1 \\
HD 190360 & 1 \\
HD 190647 & 1 \\
HD 192263 & 1 \\
HD 195019 & 1 \\
HD 196050 & 1 \\
HD 197037 & 1 \\
HD 196885 & 1 \\
BD+14 4559 & 1 \\
HD 202206 & 2 \\
HD 204313 & 2 \\
HD 204941 & 1 \\
HD 205739 & 1 \\
HD 207832 & 2 \\
HD 208487 & 1 \\
HD 209458 & 1 \\
HD 210277 & 1 \\
GJ 849 & 1 \\
HD 212301 & 1 \\
HD 213240 & 1 \\
HD 215497 & 1 \\
tau01 Gru & 1 \\
HD 216437 & 1 \\
HD 216770 & 1 \\
51 Peg & 1 \\
HD 217107 & 2 \\
HD 217786 & 1 \\
HD 218566 & 1 \\
HD 220773 & 1 \\
HD 221287 & 1 \\
HD 222155 & 1 \\
HD 222582 & 1 \\
HD 224693 & 1
\enddata
\tablenotetext{1}{Number of planets with $M_{p} \sin{i} >
0.1~M_{\mathrm{Jup}}$.}
\end{deluxetable}

\end{document}